\documentclass[nofootinbib, prd, twocolumn]{revtex4}
\usepackage{amsmath}
\usepackage{amsfonts}
\usepackage{amssymb}
\usepackage{dsfont}
\usepackage{mathrsfs}
\usepackage{bm,slashed}
\usepackage{graphicx}
\usepackage{color}
\usepackage{hyperref}
\usepackage{MnSymbol}
\hypersetup{
    colorlinks=true, 
    linktoc=page,    %set to all if you want both sections and subsections linked 
    linkcolor=blue,  
    urlcolor=black
}

\DeclareFontFamily{U}{rsfs}{}         % Formal Script            %
\DeclareFontShape{U}{rsfs}{m}{n}{<5> rsfs5 <6><7> rsfs7          %
  <8><9><10><10.95><12><14.4><17.28><20.74><24.88> rsfs10}{}     %
\DeclareMathAlphabet{\mathfs}{U}{rsfs}{m}{n}                     %
\newcommand{\be}{\nopagebreak[3]\begin{equation}}
\newcommand{\ee}{\end{equation}}

\newcommand{\bee}{\nopagebreak[3]\begin{equation*}}
\newcommand{\eee}{\end{equation*}}

\newcommand{\ba}{\nopagebreak[3]\begin{eqnarray}}
\newcommand{\ea}{\end{eqnarray}}

\newcommand{\bea}{\nopagebreak[3]\begin{eqnarray*}}
\newcommand{\eea}{\end{eqnarray*}}

\newcommand{\la}{\label}
\newcommand{\n}{\nonumber}
\newcommand{\R}{\mathbb{R}}
\newcommand{\Z}{\mathbb{Z}}

\def\pa{\partial}
\def\rd{\mathrm{d}}
\newcommand{\va}{\scriptscriptstyle}

\def\be{\begin{equation}}
\def\ee{\end{equation}}
\def\ba{\begin{eqnarray}}
\def\ea{\end{eqnarray}}

\def\label{\langle}

\begin{document}
\title{Electromagnetic duality and central charge}
\author{Laurent Freidel}
\email{lfreidel@perimeterinstitute.ca}
\author{Daniele Pranzetti}
\email{dpranzetti@perimeterinstitute.ca}
\affiliation{Perimeter Institute for Theoretical Physics,
31 Caroline St. N, N2L 2Y5, Waterloo ON, Canada}

\begin{abstract}
We provide a full realization of the electromagnetic duality at the boundary by extending the phase space of Maxwell's theory through the introduction of  edge modes and their conjugate momenta. We show how such extension, which follows from a boundary action, is necessary in order to have  well defined canonical generators of the boundary magnetic symmetries. In this way, both electric and magnetic soft modes are encoded in a boundary gauge field and its conjugate dual. This implementation of the electromagnetic duality has striking consequences. In particular, we show first how the electric charge quantization follows straightforwardly from the topological properties of the $U(1)$-bundle of the boundary dual potential. Moreover, having a well defined canonical action of the electric and magnetic symmetry generators on the phase space, we can compute their algebra and reveal the presence of a central charge between them. We conclude with possible implications of these results in the quantum theory.
\end{abstract}
\maketitle

\section{Introduction}

A fascinating property of vacuum Maxwell equations of the electromagnetic (EM) field $F$ is that they are left invariant by the duality transformation $F\rightarrow \star F$, where the $\star$ denotes the 4D Hodge dual. 
In presence of a matter source generating an electric current, this duality is broken in the bulk. It is nevertheless still preserved on boundaries away from the sources. Therefore, a boundary represents the natural place where to study the full implications of the EM duality.

The introduction of a boundary  requires the specification of boundary conditions in order for the action principle to be well defined and the symplectic potential to be conserved and closed. The usual story (see \cite{ Strominger:2017zoo} for a recent review) is that, as a consequence of this treatment, gauge invariance is  broken at the boundary, turning boundary gauge transformations into symmetry transformations  and revealing new boundary scalar degrees of freedom.  
Asymptotically the generators of the boundary symmetry are  the  
soft electric charges encoded into the boundary normal electric field.
The electromagnetic duality implies that one should be able to equip the phase space of Yang--Mills theory non only with  a notion of soft electric charge but also with a notion of soft magnetic charge, encoded into the boundary tangential magnetic field.
The main puzzle that we investigate in this work, in the context of finite boundaries, is the fact that the usual  phase space of QED inside a finite region does not allow for the possibility to define the soft magnetic charge. In particular it is not possible to define its action on the phase space variables of QED.

The resolution of this puzzle lies into a simple idea: in the presence of the boundary the phase space of gauge theories needs to be extended by  {\it edge modes}, which are conjugated to the soft electric charges. This extension is needed in order to account for the proper calculation of entanglement entropy \cite{Donnelly:2014fua, Donnelly:2016mlc}, and it is also necessary to define the fusion product allowing for the division of a gauge system into subsystems \cite{Donnelly:2016auv}, as well as  to preserve gauge invariance in the presence of boundaries \cite{Gervais:1976ec, Wadia:1976fa,Balachandran:2013wsa}. Finally, the presence of these additional boundary degrees of freedom allows one to clearly understand that boundary symmetries are not broken gauge transformations, but true symmetries associated to gauge invariant observables \cite{Donnelly:2016auv, Geiller:2017whh}. 

For completeness let us mention that the theory of edge and soft modes has a long and rich history. For the earlier instances of the appearance of new edge degrees of freedom at the boundary in gauge theories see
\cite{Gervais:1976ec, Wadia:1976fa, Kulish:1970ut,  Balachandran:2013wsa}; their role at  finite boundaries has been recently investigated in \cite{Donnelly:2016auv, Geiller:2017whh, Speranza:2017gxd}. A new understanding of their role at null infinity has been described in  \cite{Strominger:2013lka,Kapec:2015ena,Kapec:2017tkm}, while their dynamic at spatial infinity has been described in \cite{Campiglia:2017mua, Henneaux:2018gfi}.

The main point of the current paper is to show that  these additional edge modes are in fact necessary to realize EM duality and resolve the puzzle about the soft magnetic charges.
We show that the boundary extension of the phase space is exactly what is needed to define the dual magnetic charge.

This has many consequences. The first one is that it provides a clear example of the fact that  the presence of boundary soft charges does not have to be tied up with bulk gauge symmetries. The usual formulation of QED does not have a bulk notion of dual gauge transformations and still it possess boundary magnetic charges that act on the edge modes. This fact, which challenges one of the most commonly shared view about soft modes, has been sharply demonstrated recently by Campiglia et al. \cite{Campiglia:2017dpg} in the context of a massless scalar field: They have shown that there exists  soft theorems controlled by the conservation of soft charge even if the system possesses no gauge invariance.

The second consequence is the understanding  that EM duality is a  duality
between boundary charges that does not have to be realized as a bulk duality. In fact, it is known that if there is electric charge in the bulk the dual potential doesn't exist and the possibility to have a fully duality invariant formulation in the bulk disappears. What we show is that, even in that context, the EM duality is still realized as a duality between the edge modes. Since it is the edges modes that carry information about electric and magnetic charges, this is enough.

 A third consequence, which is far more reaching, 
 follows from the fact that one can now compute, in the extended phase space, the commutator algebra of electric and magnetic charges.
 Remarkably, one finds that this algebra possess a non trivial central charge. In other words, the magnetic and electric soft charges do not commute with each other.
This  fact has far reaching consequences we believe.
With hindsight, this non-commutativity of electric and magnetic charges, which goes beyond  the Dirac quantization condition, has been hinted before in several instances but never revealed explicitly.
It is related to a recent analysis of Strominger et al. \cite{Nande:2017dba} that relates the cusp anomaly to  a current algebra inside the asymptotic phase space of QED. It is also present in the work of Witten \cite{Witten:1995gf},  which has shown that nontrivial bundle summations are necessary to insure that the  compact  QED partition function transforms as a modular form under S-duality. 
Finally the closest work in spirit to our results is the work of Freed at al. \cite{Freed:2006yc} that shows that the proper understanding of Wilson lines and `t Hooft operators in QED on manifold with torsion leads to the presence of a nontrivial  Heisenberg algebra  between electric and magnetic charges.

The strategy applied here for the study of boundary degrees of freedom in Maxwell's theory follows closely the analysis done in \cite{Freidel:2016bxd}
of first order 4d gravity. There the presence of a background distributional curvature at the boundary was shown to yield a central charge in the algebra of symmetry charges associated to tangent diffeomorphisms. Our work is also related to the recent result \cite{Freidel:2017wst,Freidel:2017nhg} in string theory, where it was shown that the edge modes presents in the compact boson not only  double the dimension of the space on which the effective field theory live \cite{Aldazabal:2013sca} but also renders it non commutative.

%It would be interesting to investigate further the relationship between the two approaches. 

\section{The electromagnetism phase space} \label{PS}

Let us consider a 4D manifold $M$ with cylinder topology $M=I\times \Sigma$, where $I$ is a time interval and $\Sigma$ a space-like hypersurface. 
The Lorentzian metric on $M$ can be written as $g=-dt^2+{}^3g$, and we can choose local coordinates $x^i, i=1,2,3$  on an open subset $U\subseteq \Sigma$ and let $t$ denote the time coordinate.
 We are interested in the case where $\Sigma$ contains a boundary  $\pa \Sigma$, 
 %{\bf we have to specify what type of boundary condition we specify for the maxwell field}
  so that $\partial M=\Sigma_+ \cup \Sigma_- \cup \Delta $ where $\Sigma_\pm$ are the initial and final slices and $\Delta= \R\times \pa \Sigma$ is the time-like component of the boundary. In the following,  the space-like bulk region $\Sigma$ is assumed to have the topology of a ball and $\pa \Sigma$ the one of a 2-sphere $S^2$. The standard EM action is given by 
\be\la{SEM}
S_{EM}=\frac{1}{2e^2}\int_M F\wedge \star F + \int_M A\wedge \star J\,,
%=\frac{1}{2e^2}\int_M dA \wedge \stardA\,,
\ee
where $A$ is the EM potential, i-e a $U(1)$-valued one-form connection, $F= \rd A$ is its curvature and $\star$ denotes the 4D Hodge dual, while $J$ denotes the matter current. 
We assume in the following that the pull-back of $J$ on the boundary $\pa M$ vanishes and we do not need to specify which matter Lagrangian we are using for our analysis. From the variation of the action  we can read off  the Maxwell's equations
\be\la{Max}
\rd \star F= e^2\star J
\ee
and the pre-symplectic potential for the Maxwell's fields
\be
\Theta(\delta)= \frac{1}{e^2}\int_{\Sigma} \delta A \wedge \star F
%=\frac{1}{e^2}\int_{\Sigma} \delta A_i E^i d^3x\,,
\ee
The pre-symplectic form for the EM field reads
\be\la{O}
\Omega=\frac{1}{e^2}\int_{\Sigma} \delta A \wedge \star \rd\delta A\,.
\ee
The total symplectic potential also contains the matter symplectic potential $\Omega_m$.
From this we see that the EM phase space is parametrized by the vector potential $A_i$ on $\Sigma$ and its conjugate momentum
which is the pull back of $*F$ on the space-like slice. For a constant time surface this is simply the electric field 
\be\la{E}
E^i\equiv \partial^i A_t-\partial_t A^i\,,
\ee
with Poisson brackets
\be\la{PB}
\{A_i(x), E^j(y)\}=-e^2 \delta^j_i \delta^3(x-y)\,.
\ee

%We are now going to assume that the current  is composed of a single relativistic particle current, that is 
%\be 
%J_a(x) = q \int \rd \tau  \dot{y}_{a} \delta^4(x-y(\tau))
%\ee
%where  $q$ is the charge of the particle and $L:\tau \to y^a(\tau)$ is the  particle worldline.

\section{Electric charges}\la{sec:QE}

From the pre-symplectic form \eqref{O} we can read the Hamiltonian generator $Q^{\va E}(\alpha)$ of gauge transformations acting on $A$
\be\la{gt}
\delta_\alpha A=-\rd\alpha\,,
\ee
from its canonical properties. This generator can be decomposed into a ``soft'' part $Q^{\va E}_S(\alpha)$ and a ``hard'' component $Q^{\va E}_H(\alpha) $,
namely $Q^{\va E}(\alpha)=Q^{\va E}_S(\alpha)+ Q^{\va E}_H(\alpha)$ with
\be
\Omega(\delta_\alpha,\delta)=\delta Q^{\va E}_S(\alpha)\,,\quad
\Omega_m(\delta_\alpha,\delta)=\delta Q^{\va E}_H(\alpha)\,.
\ee
These components are given by 
 the Hamiltonian electric charge
\ba\la{QE}
Q^{\va E}_S(\alpha)&=&\frac{1}{e^2}\int_{\Sigma } \rd\alpha\wedge \star F=-\frac{1}{e^2}\int_{\Sigma }\partial_i\alpha\, E^i d^3x\,, \cr
Q^{\va E}_H(\alpha)&=&\int_{\Sigma } \alpha \star J=\int_{\Sigma }\alpha\, J^t d^3x\,.
\ea
By means of the Poisson bracket \eqref{PB}, we can immediately check that \eqref{QE} generates gauge transformations
\ba
\delta_\alpha A_i&=&\{A_i(x), \frac{1}{e^2}\int_{\Sigma }\partial_i\alpha(y) E^i(y) d^3y\}=-\partial_i\alpha\,,\n\\
\delta_\alpha E_i&=&0\,.
\ea
It is important in this derivation that the derivative terms acts on $\alpha$ and not on $E^i$, in order for the charge to be differentiable even in the presence of a boundary (The importance of this fact is clearly explained in  \cite{Balachandran:2013wsa}). 
It is also important to appreciate that the splitting between soft and hard charges is not canonical, it depends on the chosen time slice, with the exception of global charge associated with a constant $\alpha$. As times goes by, hard charges get distilled into soft charges, while the total charge is unchanged for transformations that leave the boundary of $\Sigma$ fixed.
This can be easily seen  if one uses Maxwell's equation 
(\ref{Max}). One gets that the total charge is evaluated by a boundary integral
\ba\la{QE2}
Q^{\va E}(\alpha)
%&=&\frac{1}{e^2}\int_{\Sigma } \alpha\, \rd\star F-\frac{1}{e^2}\int_{S^2 } \alpha \star F\n\\
&\hat{=}&\frac{1}{e^2}\int_{S^2} \alpha \star\! F\,,
\ea
where the hatted equality refers to the on-shell evaluation.
From this it is clear that the total charge depends only on the boundary value of $\alpha$. In particular, the total charge vanishes if $\alpha$ vanishes on $S^2$. This means that transformations (\ref{gt}) are  gauge transformations when $\alpha\overset{S^2}=0$ while they are only symmetry transformations when $\alpha$ doesn't vanish on $S^2$.
From the Poisson bracket \eqref{PB}, it is immediate to see that
\be
\{Q^{\va E}(\alpha), Q^{\va E}(\beta)\}=0\,.
\ee
$Q^{\va E}(\alpha)$ denotes the generators of an electric boundary symmetry and it is attached to a particular boundary sphere $S^2$.
The usual interpretation is then that the presence of a boundary simply breaks gauge invariance. And that the transformations  with a non vanishing parameter on the boundary are now symmetries of the system.

As understood recently \cite{Donnelly:2016auv}, this interpretation is not  satisfactory  because it omits to include in the phase space the Goldstone modes that are conjugate to the symmetry transformation.
The proper interpretation instead is as follows: The presence of a boundary $\pa\Sigma=S^2$ in a gauge theory {\it does not} break gauge invariance. Instead  it requires the {\it extension} of the phase space at the classical level and Hilbert space at the quantum level  by new degrees of freedom.
These new degrees of freedom form a boundary canonical pair $(\varphi,\pi)$ that lives on $S^2$. $\varphi$ is  the {\it edge mode} and $\pi$ is its canonical momentum.
In practice, this means that the symplectic potential is given by
\be\la{ext-pot}
\Theta_{ext}= \frac{1}{e^2}\int_{\Sigma} \delta A \wedge \star F
+\int_{S^2} \delta \varphi \pi.
\ee

These edge modes are physical degrees of freedom that form a canonical pair located  at the boundary of the physical region. They represent non local degrees of freedom that are revealed by boundaries or non trivial topology (see \cite{Rovelli:2013fga} for an enlightening and elementary account of this phenomenon). 
In this extended phase space  we can identify a generator of gauge transformation
$(A,\varphi,\pi)\to (A+\rd \epsilon,\varphi-\epsilon, \pi)$ that contains a bulk component and a boundary component. 
It is explicitly given by
\be
G(\epsilon)= \frac{1}{e^2} \int_\Sigma \rd \epsilon \wedge \star F + \int_\Sigma \epsilon  \star J-  \int_{S^2} \epsilon \pi
\ee
and it satisfies the on-shell condition $ G(\epsilon)\hat{=}0$. 
The bulk component of this equation is the Gauss law
$\rd (\!\star \! F)\overset{\Sigma}= e^2\!\star \!J$  in QED and the boundary component identifies the momenta $\pi$ in terms of  the bulk fields:
\be
\star F\overset{S^2}= e^2 \pi\,,
\ee
where the symbol $\overset{S^2}=$ means that the equality is taken as an equality of forms pulled back on  $S^2$ by the  sphere embedding\footnote{In other words it means that $\imath_{S^2}^*(\star F)= e^2\pi$.} $\imath_{S^2} :S^2 \to \Sigma$.
In coordinates this simply means that $\pi  \overset{S^2}=\sqrt{h} E^r$ is the densitized radial electric field where $h$ is the metric on $S^2$ and the boundary is assumed to be at $r=cst$. 
The boundary Gauss law implies that the edge mode momentum $\pi$  is entirely determined by the boundary value of the bulk field\footnote{In the gravity context, the analog of the boundary Gauss law is given by the simplicity constraint, relating as well the value of the boundary degrees of freedom to the pullback of the bulk phase space momentum \cite{Freidel:2016bxd}.}. 
Morever  the bulk dynamic implies the conservation of the boundary current: $\rd \pi =0$.
There exists also a  boundary symmetry generator that leaves the bulk variable $A$ fixed and transforms the boundary variables in the following manner: $(\varphi,\pi)\to(\varphi+ \alpha, \pi)$.
This symmetry generator  is  simply given by
\be
Q^{\va E}(\alpha):= \int_{S^2} \alpha \pi \,. 
\ee
%In the second equality we have expressed the total charges 
On the other hand, the edge mode $\varphi$ is  a new 
physical degree of freedom  \cite{Donnelly:2016auv} that is revealed by the presence of the boundary, and it represents 
the parameter conjugate to this symmetry transformation,
i-e the analog of the Goldstone mode \cite{Strominger:2017zoo}.  Its presence allows us to define on the boundary a gauge invariant field 
\be\label{bdya}
a :\overset{S^2} = A+\rd \varphi. 
\ee 
 It is important to realize that one can  chose appropriate boundary conditions and give appropriate dynamics to the edge modes, in order  to insure that not only the global charge but also the soft charges are conserved in time. This is established in \cite{Donnelly2} (see also \cite{Blommaert:2018oue} for a study of edge dynamics in gauge theories  using the path integral).

%Using the Gauss Law the total charge can be written  as a bulk integral, which reveals that it is given by  the sum of 
% soft charges plus  hard charges involving the current:
% \be
%Q^{\va E}(\alpha)\hat{=}\, 
%\frac{1}{e^2} \int_\Sigma \rd \alpha \wedge\star F + \int_\Sigma \alpha  \starJ. 
%\ee

\section{Magnetic charges}\la{sec:QM}

It turns out that the electric symmetry generated by $Q^{\va E}(\alpha)$ which measures the boundary electric field is not the only  symmetry associated with the presence of a boundary.  There is also a corresponding magnetic symmetry generated by $Q^{\va M}(\tilde\alpha)$ which probes the boundary magnetic field (see \cite{Strominger:2015bla, Campiglia:2016hvg}  for a previous analysis of magnetic charges in the context of the soft photon theorem and its relation to  large gauge transformations
at null infinity).
Since in electromagnetism we do not allow for the presence of magnetic monopoles, we expect the magnetic charge $ Q^{\va M}(\tilde\alpha)$ to be entirely soft and given by
\be\label{QMsoft}
 Q^{\va M}_S(\tilde\alpha) \hat{=} \frac1{2\pi} \int_{\Sigma} \rd\tilde{\alpha} \wedge F.
\ee 
This doubling of the boundary symmetry group is a very important element of Maxwell theory, which is also happening in non-Abelian Yang--Mills theory, as well as in gravity, as the twistor formulation \cite{Mason:2008jy} or the Ashtekar formulation \cite{Ashtekar:1986yd} of the theory can reveal.
This is the phenomenon that we want to investigate further.
This doubling of the charge algebra is puzzling at first sight since $Q^{\va E}$ naively appeared as the boundary generator associated with gauge transformations.
It turns out that there  is no notion of magnetic gauge transformation available in electrodynamics that acts on the bulk phase space variables $(A,E)$, and therefore no notion of boundary dual gauge symmetry, if one keeps the electromagnetic phase space unextended. 
This is a subtle point related to the fact that the charge $Q^{\va M}_S(\tilde\alpha)$ is not differentiable, i-e 
\be
\delta Q^{\va M}_S(\tilde\alpha)=-\frac1{2\pi} 
\int_{S^2} \rd \tilde\alpha \wedge \delta A
\ee
depends on the boundary value of $\delta A$ and, as such, it cannot be used as a canonical generator of transformations, since this boundary value does not possess, in the unextended phase space,  a canonical conjugate. As it will be shown in Sec. \ref{sec:m-canonical}, the expression \eqref{QMsoft} is indeed not the right form of the magnetic charge derived from a well defined canonical analysis.

The purpose of our work is to investigate these puzzles. In particular, we want to understand:  What boundary symmetry group these magnetic charges represent? And what is their commutation relations with the electric charges? And eventually what is their dynamics?

The experience from the electric case clearly suggests that one also needs to have an extra canonical pair $(\tilde\varphi, \tilde\pi)$  representing the magnetic soft modes. 
The magnetic version of the Gauss law then possess a bulk and a boundary components
\be\label{bdytpi}
\rd F =0,\qquad 
\frac1{2\pi} F\overset{S^2}= \tilde{\pi}. 
\ee
These two equations are similar to their electric counterpart.
However, there are two key differences: First, the magnetic analog of the Gauss law is the Bianchi identity, i-e it is a constraint which is  identically satisfied, instead of being an equation of motion; second, the soft mode momentum is not any arbitrary 2-form but one derived from a boundary connection.
In other words, we have that $\tilde\pi = \rd a$, where $a$ is the boundary gauge field introduced in (\ref{bdya}). 

This extension can explain the possibility to have 
 a magnetic soft symmetry even if it doesn't correspond to any gauge invariance. As we are about to see, it  also allows us to introduce a conjugate variable to the boundary value of the  gauge field.  
 The challenge is then to understand how these magnetic soft modes enter the extended electric phase space and what is the action of the electric charge on the magnetic soft modes.
 We now show that in fact the electrically extended phase space already carries enough information to include the magnetic soft charges.

 \section{Electromagnetic duality and charge quantization}\la{sec:cq}

In order to understand these issues, let us assume that there are no matter currents inside $\Sigma$. In this case, we can use the fact that  vacuum Maxwell  equations are preserved by the duality  between the electric and the magnetic fields. 
This duality can be conveniently expressed by the demand that there exists a dual vector potential $\tilde A$, with curvature $\tilde{F}= \rd \tilde{A}$ and such that 
\be\la{Ftilde}
\tilde F = \frac{2\pi}{e^2}  \star F\,.
\ee
In terms of the dual potential, the vacuum Maxwell equations read
\ba
&&\rd \tilde F= 0\,,\la{dual-Max1}\\
&&\rd \star \tilde F=0\,.\la{dual-Max2}
\ea
What is remarkable about this duality is the fact that demanding its validity on the boundary sphere
implies the quantization of charge. In fact,
\be\la{q}
q := \frac{1}{e^2}\int_{S^2}  \star F =   \frac1{2\pi} \int_{S^2}\tilde F
\in \mathbb{Z}.
\ee
The result that $q$ is an integer follows from the fact that the 
last integral measures the Chern class of the dual bundle associated with $\tilde{A}$. The presence of charges hence means that the dual bundle is non trivial. We can show this with a simple example.

In presence of a source term, the equation \eqref{dual-Max1} above becomes
\be\la{dual-Max3}
\rd \tilde F= 2\pi \star J
\ee
and, thus, the dual curvature $\tilde F$ is not exact, namely its integral over certain 2-spheres is nonzero. 
%Notice that we are not introducing a magnetic monopole as a source of the dual Maxwell equations. The source of the dual eq. \eqref{dual-Max3} is still an electric current.
Let us consider the case where the unit 2-sphere $S^2$ contains a point-like electric charge $q$ at the origin $\{0\}$. In order to have non-singular solutions, we remove the origin so that our bulk becomes $\Sigma\backslash \{0\}$. In this case, the Maxwell eq. \eqref{dual-Max3} implies
\be\la{flux}
\frac{1}{2\pi}\int_{S^2 } \tilde F=  q\,.
\ee
Therefore, a point-like electric charge plays the role of a monopole for the dual potential. A solution to the previous equation is given by
\be
\tilde F=\frac{ q}{4\pi}\sin\theta\, \rd\theta\wedge \rd\phi\,.
\ee
As pointed out above, the condition \eqref{flux} implies that $\tilde F$, while closed on $\Sigma\backslash q$,  cannot  be exact everywhere there. This means that the dual potential $\tilde A$ cannot be globally defined there. We have two possibilities. We can consider the region $N=\Sigma\backslash\{z-{\rm axis}\leq 0\}$ and the dual potential
\be
\tilde A_{\va N}=\frac{ q}{4\pi}(1-\cos\theta) \rd\phi\,,
\ee
or the region $S=\Sigma\backslash\{z-{\rm axis}\geq 0\}$ and
\be
\tilde A_{\va S}=-\frac{ q}{4\pi}(1+\cos\theta) \rd\phi\,.
\ee
The two expressions for the dual potential do not agree in $N\cap S=\Sigma\backslash\{z-{\rm axis}\}$. The transition function $\tilde \alpha_{\va NS}$ in the overlap such that 
\be
\tilde A_{\va N}= \tilde A_{\va S}+\rd \tilde\alpha_{\va NS}
\ee
 is given by 
 \be
\tilde  \alpha_{\va NS}=\frac{ q}{2\pi} \phi\,.
 \ee
The fact that $\tilde \alpha_{\va NS}$ is not a single-valued function on $N\cap S$ is a reflection of the fact that the overlap region is not simply connected. Despite the transition function $\tilde \alpha_{\va NS}$ not being single-valued, it is possible to construct a complex line bundle over $S^2$ if the analog of the Dirac quantization condition is satisfied, namely if  
\be
 q \in \mathbb{N}.
\ee
 This quantization condition reflects the non-triviality of the dual $U(1)$-bundle, as a consequence of the non-triviality of the second homology group $H_2(\Sigma\backslash q,\Z)$. Therefore, while we can chose the trivial $U(1)$-bundle for the vector potential $A$, EM duality implies that for the dual potential this is not possible and the corresponding bundle is non-trivial.
 
  We want to stress out that, while a similar topological explanation of magnetic charge quantization is well known in the literature (see, e.g., \cite{Quiros:1981gd}), our derivation of the quantization of the electric charge resulting from \eqref{q} does not require the introduction of a magnetic monopole. It simply follows from a dual description of Maxwell  theory in the more general case of a singular dual gauge parameter, as further justified by our canonical analysis of magnetic charges in Sect.  \ref{sec:m-canonical}.
 
 \section{Resolution and Boundary action}\la{sec:boundA}

 In order to understand the canonical meaning of the magnetic charge $Q^{\va M}(\tilde{\alpha})$, we need to express its transformation on the phase space variables. 
This is subtle since, on the one hand, we expect the magnetic charge to be a boundary transformation of the dual potential $\tilde{A}\to \tilde{A}+\rd \tilde\alpha$, but, on the other hand, 
we also have that  both $\star F$ and $A$ are invariant under dual gauge transformations.  
In order to analyze this we need to  express the electromagnetic duality at the canonical level in the presence of boundaries.

 In the presence of boundaries,
the action (\ref{SEM}) is not differentiable unless we either restrict the variation to satisfy $\delta A\overset{\Delta}=0$ or we impose that $\star F\overset{\Delta}=0$. If we impose the former condition,  we cannot define the canonical variation of $Q^{\va M}(\tilde{\alpha})$, since it depends on the value of $A$ along the boundary; while the latter condition implies that $Q^{\va E}(\alpha)$ is superselected to vanish. In both case we cannot compute the commutator.

 We thus need to add a boundary action in order to render the action differentiable under less stringent conditions. We will see in a moment that the introduction of a boundary term in the EM action leads to an extended phase space in strict relationship with \eqref{ext-pot}, i-e capturing exactly the new degrees of freedom encoded by the edge mode $\varphi$ and its canonical momentum $\pi$ introduced in Section \ref{sec:QE}.

 Motivated by the electromagnetic duality, we introduce a  boundary gauge field $\tilde{a}$, living on $\Delta$, and consider the extended action \footnote{The boundary term in \eqref{SEM2} has been previously considered in \cite{Seiberg:2016gmd} as a coupling term for the 2+1 gauge field restricted to the boundary, once the bulk theory is re-expressed in terms of the  dual gauge field.}
\be\la{SEM2}
S'_{EM}=\frac{1}{2e^2}\int_M F\wedge \star F+ \int_M A\wedge \star J -\frac{1}{2\pi}\int_\Delta A \wedge \rd\tilde a\,.
\ee

%while $a$ is related to the bulk field and edge mode by 
%\be
%a:\overset{\Delta} = A+\rd \varphi.
%\ee
The variation of this action under $A$ imposes the same bulk EOM: $\rd\star F = e^2\star J$. It also imposes a boundary equation which identifies the value of the boundary electric field with the dual curvature 
\be\la{BC}
\star F\overset{\Delta}= \frac{e^2}{2\pi} \rd\tilde a ,
%\qquad 
%\delta \tilde F\overset{\Delta}= 0 .
\ee
where the subscript $\Delta$ means that the equality is valid for forms which  are pulled-back on the boundary. 

This  condition, which  follows from the variation of the boundary value of $A$, says that the dual curvature is proportional to the soft momenta $\pi = \frac1{2\pi} \rd \tilde{a}$, in perfect analogy with (\ref{bdytpi}).
It also means  that the potential $\tilde{a}$ entering the boundary action is, up to a dual gauge transformation,
equal to the boundary value of the  dual potential: $\tilde{a}= \tilde{A}+\rd\tilde\varphi$. 
Importantly, it doesn't require the dual potential $\tilde{A}$ to exist in the bulk, but it is enough that it exists on the boundary $\Delta$ only.
This is satisfactory since we have seen that the dual potential can only be defined away from the location of electric charges.
The boundary EOM \eqref{BC} also means that  
\be
\rd \star F(A)\overset{\Delta}= 0, 
\ee
which follows from the bulk EOM provided the component of the current $J^r$ normal to $\Delta$ vanishes on the boundary, i-e provided that no charge is crossing the boundary. 

Finally, variation of the boundary dual potential yields the boundary EOM
\be
\rd A \overset{\Delta}= 0. \la{dA}
\ee

\section{Symmetry and canonical structure}\la{sec:m-canonical}

The symplectic form associated with the action $S'$ is given by 
\be\la{Omega}
\Omega'= \frac{1}{e^2}\int_{\Sigma} \delta A \wedge \star\delta F
+ \frac{1}{2\pi} \int_{S^2} \delta a \wedge \delta\tilde{a} \,,
\ee
where we have introduced the boundary field $a \overset{\Delta}= A +\rd\varphi$ (see eq.(\ref{bdya})),
which includes the electric soft mode. The inclusion of this new boundary scalar fields relating  the bulk and the boundary gauge potentials is crucial to obtain a well defined boundary phase space structure, allowing us to derive non-vanishing soft magnetic charges, as shown in a moment.
The last term in \eqref{Omega} is similar to a  Chern-Simons symplectic term for a pair of $U(1)$ connection. Therefore
the  presence of the boundary term  yields the boundary Poisson bracket
\be\la{PBbound}
\{a_i(x), \tilde a_j(y)\}=2\pi \epsilon_{ij}\delta^2(x-y)\,.
\ee

This shows that in addition to the electric soft pair $(\varphi,\pi=\rd \tilde{a})$, the dual connection $\tilde{a}$ is the variable conjugated to the boundary gauge potential $A$ pulled back on $S^2$. This is the variable needed in order to be able to define the action of the magnetic charge on the phase space.
In this formulation, the gauge transformation acts as $\delta_\epsilon(A,a,\tilde{a})=(\rd \epsilon,-\rd\epsilon,0)$,
while the electric  symmetry is acting only on the electric boundary fields $\Delta_\alpha(A,a,\tilde{a})=(0,\rd\alpha,0)$.
The magnetic symmetry is also acting only the boundary  fields as
$\tilde\Delta_{\tilde\alpha}(A,a,\tilde{a})=(0,0,\rd\tilde\alpha)$.
All these symmetry have now canonical generators
\ba
 \Omega(\delta_\epsilon,\delta)&=&\delta G(\epsilon)\,,\\
 \Omega(\Delta_\alpha,\delta)& =& \delta Q^{\va E}(\alpha)\,,\\
  \Omega(\tilde\Delta_{\tilde\alpha},\delta) &=& \delta Q^{\va M}(\tilde\alpha)\,.
\ea
The Gauss law and the electric symmetry generators are the same as before, namely
\ba
G(\epsilon)
&=& \frac{1}{e^2}\int_{\Sigma} \rd \epsilon  \wedge\star F 
+ \frac{1}{2\pi} \int_{S^2}  \rd \epsilon \wedge \tilde{a} \hat{=} 0\,,\\
Q^{\va E}(\alpha)&=&   
 \frac{1}{2\pi} \int_{S^2}   \rd \alpha \wedge  \tilde{a}\,,\la{QEc}
\ea
while   the magnetic charge is given by  the following expression 
\be\la{QMc}
Q^{\va M}(\tilde \alpha):=
%-\frac{1}{e^2}\int_{S^2 }\tilde \alpha\, F=
\frac{1}{2\pi}\int_{S^2 } \rd\tilde\alpha\wedge a\,.
%=\frac{1}{e^2}\int_{S^2 } \partial_i\tilde\alpha\, A_j \epsilon^{ij}d^2x \,.
\ee

The important thing to observe about the expression \eqref{QMc} is that, unlike the electric charge case, we no longer need $\tilde{\alpha}$ to be a well defined scalar on $S^2$; we only need its exterior derivative to be well-defined on $S^2$.
More generally, the magnetic charge depends only on a closed one-form $\tilde\omega= \rd \tilde\alpha$ which is locally but not globally exact.  In particular, we can consider the case where $\tilde \alpha$ introduces a branch cut so that  $\rd \tilde\alpha$ admits  poles on the boundary 2-sphere. For instance,  given the spherical coordinates $(\theta,\phi)$ on the sphere, the form $\rd \tilde \alpha=\rd\phi$ introduces a singularity at the north and south poles. The presence of these singularity is tied up to the fact that the dual bundle is in general non-trivial in the presence of electric charge, as we saw in Section \ref{sec:cq}. The location of these singularities represents the possibility of dual electric monopoles. Despite the singularity of the gauge parameter, the magnetic charge is still well defined.
In this more general case, the surface $S^2$ is no longer compact but contains circle boundaries around the poles. Therefore, when there is a finite set of poles  $\{p\}$, we can rewrite the magnetic charge as
\be\la{QM2}
Q^{\va M}(\tilde \alpha)=-\frac{1}{2\pi} \int_{S^2\backslash \{ p \}} \tilde \alpha\, F+\frac{1}{2\pi}\sum_p\oint_p \tilde\alpha a\,.
\ee
This shows clearly how the form of the well defined magnetic charge \eqref{QMc} is {\it not} equivalent to the  guess \eqref{QMsoft} once we allow for the more general case of the dual gauge parameter to have singularities. In particular, the boundary EOM \eqref{dA} does not imply the vanishing of the magnetic charge  \eqref{QMc}, but it simply allows us to rewrite the charge as a circle integral around the singularities. 
We are going to elucidate the physical implications of allowing for this more general set of gauge parameters in the next Section.

Let us first point out that another way to reveal the duality symmetry of the extended phase space is to write the symplectic form \eqref{Omega} as
%\ba
%\Omega&=&\frac{1}{e^2}\int_{\Sigma} \delta A \wedge \star \delta F=
%\frac{1}{2\pi}\int_{\Sigma} \delta A \wedge  \rd\delta \tilde A\n\\
%%&=&\frac{1}{e^2}\int_{N}  \rd\delta A \wedge \delta \tilde A+\frac{1}{e^2}\int_{S^2_{\va N}} \delta A \wedge \delta \tilde A\n\\
%&=&
%-\frac{1}{2\pi}\int_{\Sigma}   \delta\tilde A\wedge  \delta F   +\frac{1}{2\pi}\int_{S^2} \delta \tilde{a} \wedge \delta   a\,\cr
%&=&\frac{e^2}{(2\pi)^2}\int_{\Sigma}    \delta\tilde A\wedge  \star\delta \tilde{F} +\frac{1}{2\pi}\int_{S^2} \delta \tilde{a} \wedge \delta   a
%\ea
\bea
\Omega' &=& \frac{1}{2\pi }\int_{\Sigma} \delta A \wedge \delta\tilde{F}
+ \frac{1}{2\pi} \int_{S^2} \delta a \wedge \delta\tilde{a} \cr
&=& \frac{1}{2\pi }\int_{\Sigma} \delta F \wedge \delta\tilde{A}
+ \frac{1}{2\pi} \int_{S^2} \delta a \wedge \delta(\tilde{a}-\tilde{A})\cr
&=&\frac{e^2}{(2\pi)^2}\int_{\Sigma} \delta\tilde{A} \wedge \star\delta \tilde{F} 
+\int_{S^2}  \delta \tilde \pi \, \delta\tilde{\varphi} .
\eea
where we have used in the first line that $*F=\tilde{F}$, then integrated by part and used in the last equality that $\tilde\pi = \tfrac1{2\pi} \rd {a}$ while $\tilde{a}-\tilde{A} =\rd \tilde\varphi$.
This expression  for the symplectic form shows that, under the duality transformation, the canonically conjugate  bulk variables are the dual potential and dual electric field $(\tilde{A},\star \tilde F)$, with the charge $e$  replaced by the dual charge $2\pi/e$. The boundary phase space is also  parametrized by the dual variables, represented by the magnetic edge mode and its conjugate momentum. This represents the exact dual of the extended phase space \eqref{ext-pot} introduced in Section \ref{PS}.

\section{Charges algebra}

Like its electric version, the magnetic charges commute 
\be 
\{Q^{\va M}(\tilde{\alpha}),Q^{\va M}(\tilde{\beta})\}=0.
\ee

However, one of the main consequences of allowing for the dual monopole singularities  for the dual gauge transformations generated by the magnetic charges is   the appearance of a central charge  in the algebra of the mixed sector of EM, as shown in a moment. And second, the magnetic transformation generated by $Q^{\va M}(\tilde{\alpha})$ naturally appears in the dual formulation of QED in the presence of electric charges.

Let us now derive another striking implication of the EM duality, in addition to the electric charge quantization obtained in Section \ref{sec:cq}, by computing the  algebra between electric and magnetic charges. We saw above that, in order to fully accomodate the EM duality at the boundary, we need to extend the phase space. This allowed us to write the canonical generators of the electric and magnetic symmetry as the boundary integrals \eqref{QEc}, \eqref{QMc}. We can thus use the boundary Poisson bracket \eqref{PBbound},
derived from the extended symplectic form  \eqref{Omega}, to compute their algebra. An immediate calculation yields
\ba
\{Q^{\va M}(\tilde\alpha), Q^{\va E}(\alpha)\}
&=&\frac{1}{2\pi }  \int_{S^2} \rd \tilde \alpha\wedge \rd \alpha\n\\
&=&-\frac{1}{2\pi} \sum_p \oint_p \alpha\, \rd\tilde\alpha\,.
\ea
This shows that the boundary symmetry algebra of electromagnetism possesses a central charge\footnote{A paper \cite{Hosseinzadeh:2018dkh} appeared while we were completing our work, which also  proposed a nontrivial algebra between soft  electric and magnetic charges. The approach there still relies  on the early  premise that boundary symmetry are broken gauge. It therefore necessitates a non Lagrangian formulation with an extra duality constraint which  implements  by hand the EM duality in the bulk. As we already mentioned, the bulk duality is not  satisfied if charges are present, while the boundary duality survives. }%Let us also clarify that, we had pointed out to one of the authors the appearance of a central charge when taking the magnetic sector into account. }. 

The appearance of a central charge on circle boundaries around the poles is natural once we notice that the boundary term in the symplectic form \eqref{Omega} corresponds to the one of a $U(1)\times U(1) $ Chern-Simons theory, and this is well known to lead to a $U(1)\times U(1) $ Kac-Moody algebra on a circle when one considers observables associated to global symmetry transformations \cite{Balachandran:1991dw}.

\section{Quantum implications}

The EM duality involves a doubling of the  boundary symmetry, but the presence of this central charge implies that  at the quantum level we cannot select a vacuum state invariant under both symmetries: If we select the magnetic vacuum, then an infinite tower of electric edge modes arises and vice-versa. In either case, we  end up with a series of inequivalent EM vacua parametrized by the Goldstone soft boson of this spontaneously broken EM duality symmetry. The non trivial commutation relation between electric and magnetic charge shows that in effect the electric soft mode conjugated to the electric soft charge can be understood as the dual magnetic charge. 

In QED, and in the regime where one can define the $S$-matrix,  it is customary to set all the asymptotic soft magnetic charges to zero. This follows from the fact that it is impossible to create non trivial asymptotic soft magnetic charges from classical sources that have  compact support \cite{Winicour:2014ska}. This leads to the belief that magnetic charges cannot be excited. Here we see that in fact this is simply an artifact of the restriction to classical configuration.  Our derivation shows that we do not need magnetic monopoles in order to produce a state with  non trivial magnetic soft charge, we  need superposition. Since electric and magnetic charges are conjugated in some ways a state diagonalising the soft magnetic charge  would simply correspond to a superposition of electric sectors and it is therefore a purely quantum phenomenon\footnote{Let us point out that the inclusion of magnetic monopoles in our analysis would allow us to extend the EM duality in the bulk,  leading to the appearance of a hard component for the magnetic charge \eqref{QMc},  but it would not modify the main results of this manuscript; only the interpretation of some of their physical implications would be different. For instance, the existence of magnetic monopoles would allow for soft magnetic charges already at the classical level.}. 

It is important to appreciate that the possibility of having quantum  superposition  of electric charges resulting in an effective magnetic charge is a non-perturbative effect that relies on the relationship 
$ \tilde F = \frac{2\pi}{e^2}  \star F$. In the perturbative limit $e\to 0$ the duality is lost and the QED field effectively becomes a non-compact gauge field. There is a system in 2d which is similar to the 4d QED. It is the compact massless boson $\varphi \sim \varphi + R$. For this system the duality relation is 
$\rd \tilde\varphi =\frac{2\pi \alpha' }{R^2} *\rd \varphi$  (with $\alpha'$ encoding a length scale) and the 2d analog of the electric charge is played by the radius of compactification. It has recently been  shown that the zero mode of the compact boson  and its dual are non commuting \cite{Freidel:2017nhg,Freidel:2017wst}, a phenomenon which is a precursor to the non-Abelian charge algebra we just found. What this suggests is then that the theory of QED in which the duality of charge is implemented is in fact a theory of compact QED, with the charge playing the role of the radius of compactification.
It is expected  that, on the lattice, compact  QED is confining \cite{Polyakov:1975rs,Polyakov:1987ez}, due to monopole condensation that allows for the transition between an electric vacuum with $Q^{\va M}=0$ to a confining vacuum with $Q^{\va E}=0$. Therefore, this suggests that the phenomena we just described  can for the first time allow us to define a compact QED theory in the continuum and open up the study of monopole condensation outside of the lattice regularisation.

Finally, we hope our study shed new light on the nature of the soft modes. It is now well established \cite{Kulish:1970ut, Kapec:2015ena, Kapec:2017tkm}  that the QED vacuum state is degenerate under the action of the electric symmetry charges \eqref{QEc}. The conjugate variable controls the  creation of soft photon modes, as we saw in Section \ref{sec:QE}, and it can be understood  in terms of wavefunction dressing operators for outgoing particles in the $S$-matrix elements of QED, in  order to have IR finite scattering amplitudes \cite{Kapec:2017tkm}. Our results connect this dressing with soft magnetic charges. It becomes therefore of crucial importance to propose experimental detection of the soft modes entering the dressing. One way is to focus on the measure of  electromagnetic memory of the vacuum state, encoded in the 
different gauges of the potential field
 as soft photon clouds generated in scattering processes cross the boundary,
 through the phases of test particles in the Aharonov--Bohm effect.
%In fact, similarly to the magnetic charge case \eqref{QM2}, in presence of poles, we can write the electric charge as
%\be\la{QE2}
%Q^{\va E}( \alpha)=-\frac{1}{2\pi} \int_{S^2\backslash \{ p \}}  \alpha\, \tilde F+\frac{1}{2\pi}\sum_p\oint_p \alpha \tilde a\,.
%\ee
A suggestion along these lines was, for instance, made in \cite{Susskind:2015hpa, Hamada:2017bgi}.

%\section{Rest}

%
%This duality can be expressed in terms of the electric and magnetic soft momenta by the fact that there exist potential $a$ and $\tilde{A}$ such that
%\be
%\pi =\rd \tilde{a},\qquad \tilde\pi =\rd a. 
%\ee
%In terms of the dual potential, the symplectic potential read
%\be
%\Theta_{ext}= \frac{1}{e^2}\int_{\Sigma} \delta A \wedge\star F
%- \int_{S^2} a \wedge \delta\tilde{a} +\oint_C  \delta \varphi   \tilde{a} .
%\ee
%$ a\overset{S^2}= A-\rd \varphi$
%\be
%\Theta_{ext}= \frac{1}{e^2}\int_{\Sigma} \delta A \wedge\star F
%+ \int_{S^2} \delta  (a -A) \tilde{a} +\oint_C  \delta \varphi   \tilde{a} .
%\ee
%\be
% \delta \tilde{A} \curlywedge \delta A- \delta \tilde{a} \curlywedge \delta  (a -A) =  \delta  (\tilde{a} -\tilde{A}) \delta {a}
%\ee
%\be
%\Theta_{ext}= \frac{1}{e^2}\int_{\Sigma} \delta A \wedge\star F
%+ \int_{S^2} \delta  (a -A) (\tilde{a}-\tilde{A}) +  
%\int_{S^2} \delta  (a -A) \tilde{A}  .
%\ee
%

\acknowledgments

 Research at Perimeter Institute for Theoretical Physics is supported in part by the Government of Canada through NSERC and by the Province of Ontario through MRI. LF would like to thank Lee Smolin and Djordje Minic and Rob Leigh for friendly support and critical inputs.
 We would like to acknowledge intersting discussions about EM duality with W. Donnelly and M. Geiller.
 LF would also like to acknowledge a very enlightening discussion with A. Strominger about central charges  and the importance of the magnetic sector that was critical for the formulation of the idea presented here.

\bibliographystyle{uiuchept}
 
%\bibliography{Soft-modes}

% \bibliographystyle{apsrev-title}

\providecommand{\href}[2]{#2}\begingroup\raggedright\endgroup

\end{document}